# A collaborative theoretical and experimental study of the structure and electronic excitation spectrum of the BAr and BAr$_2$ complexes

Millard H. Alexander, Andrew Walton, and Moonbong Yang

*Department of Chemistry and Biochemistry, University of Maryland*

*College Park, MD 20742-2021*

Xin Yang, Eunsook Hwang,[a] and Paul J. Dagdigian

*Department of Chemistry, The Johns Hopkins University*

*Baltimore, MD 21218-2685*

**Abstract**

We report the investigation of the $3s \leftarrow 2p$ transition in the BAr$_2$ cluster. In a supersonic expansion of B atoms entrained in Ar, at high beam source backing pressures we observe several features in the fluorescence excitation spectrum which cannot be assigned to the BAr diatom. Using BAr($X$, $B$) potential energy curves which reproduce our experimental observations on this molecule and an Ar–Ar interaction potential, we employ a pairwise additive model, along with variational and diffusion Monte-Carlo treatments of the nuclear motion, to determine the lowest vibrational state of the BAr$_2$ cluster. A subsequent simulation of the fluorescence excitation spectrum reproduces nearly quantitatively the strongest feature in our experimental spectrum not assignable to BAr. Because of the barrier in the BAr($B^2\Sigma^+$) potential energy curve, the $3s \leftarrow 2p$ transition in the BAr$_2$ is predicted to have an asymmetric profile, as is found experimentally.

[a] Present address: Molecular Physics Laboratory, SRI International, Menlo Park, CA 94025.

I. INTRODUCTION

There has been considerable interest in non-bonding interactions between metal and rare gas atoms. In part, this interest stems from a desire to understand the behavior of light metal atoms in cryogenic media, such as solid rare gas and hydrogen matrices.[1] Such systems offer the potential for the storage of energy for technological applications. Spectroscopic investigations of electronic transitions in diatomic metal – rare gas van der Waals complexes have provided considerable information on such interactions in both ground and electronically excited states.[2] The availability of these diatomic potential energy curves is essential for understanding the behavior of metal atoms in larger clusters and cryogenic matrices.

We have been interested for some time in the non-bonding interactions of the boron atom with rare gases and the hydrogen molecule.[3-14] There are two B atomic transitions with wavelengths > 200 nm: $2s^23s\ ^2S \leftarrow 2s^22p\ ^2P$ at 249.8 nm and $2s2p^2\ ^2D \leftarrow 2s^22p\ ^2P$ at 208.9 nm.[15] With laser fluorescence excitation detection, we have observed diatomic electronic transitions correlating with both of these transitions in the BNe and BAr complexes. The lowest electronic transition, $B^2\Sigma^+ - X^2\Pi$, whose upper and lower states correlate with the B atomic $2s^23s\ ^2S$ and $2s^22p\ ^2P$ states, respectively, has a dramatically different spectrum for these two species. For BNe, this transition appears as a broad, unstructured feature to the blue of the $2s^23s\ ^2S \leftarrow 2s^22p\ ^2P$ atomic transition.[5] With the help of electronic structure calculations, this spectrum was simulated, and the unstructured nature of the transition was shown to be due to the purely repulsive nature of the BNe($B^2\Sigma^+$) potential energy curve.[5] The spectrum of the BAr $B^2\Sigma^+ - X^2\Pi$ transition displays rotationally resolved bands in the red portion of the spectrum and several unresolved features to the blue. The potential energy curve for the BAr($B^2\Sigma^+$) possesses a deep inner well [$D_0' \approx 1010$ cm$^{-1}$] and a shallow outer well, with an intervening barrier.[3] The features in the blue portion of the spectrum are not rotationally resolved because of the large tunneling rate through and over this barrier.

There has been considerable interest in the effect of solvation on electronic transitions of a chromophore.[16, 17] Electronic absorption spectra of boron atoms trapped in argon and hydrogen matrices have been reported.[1, 18, 19] One surprising observation from these studies is that the electronic absorptions appear only at wavelengths < 220 nm, considerably shorter in wavelength than that of the first transition in atomic B, $2s^23s\ ^2S \leftarrow$ $2s^2 2p\ ^2P$ at 249.8 nm. It is possible that this gas-phase transition is shifted considerably to the blue,[18] or so broadened by interaction with the matrix as to be undetectably weak and blended into the baseline absorption. Boatz and Fajardo[20] have carried out classical Monte Carlo simulations of the B $3s \leftarrow 2p$ absorption in a cluster of argon atoms and conclude that the transition should remain sharp in such an environment.

One approach to the understanding of the spectrum of a chromophore in a matrix is to study the transition in small clusters. Several groups have studied the Hg $6s6p\ ^3P_1 \leftarrow 6s^2\ ^1S$ transition in small $HgAr_n$ clusters.[21-23] Visticot *et al.* have studied the Ba $6s6p\ ^1P \leftarrow 6s^2\ ^1S$ transition for Ba atoms attached to large $Ar_n$ clusters.[24] Whetten and co-workers[25] have probed the evolution of the Al $4s \leftarrow 3p$ transition in $AlAr_n$ clusters as a function of the size $n$. More recently, Okumura and co-workers[26] have carried out similar experiments on the $3d \leftarrow 3p$ transition in $AlAr_n$. Scoles and co-workers have investigated electronic transitions of Na atoms and dimers attached to liquid helium clusters.[27]

Theoretical investigations of the absorption spectrum of alkali metal atoms in noble gas solids and/or clusters have been reported by Singer and co-workers,[28] and Boatz and Fajardo.[29] Cheng and Whaley have carried out a similar investigation involving Li atoms in $p$-$H_2$ clusters.[30] Zúñiga *et al.* have reported the theoretical study, of the absorption spectrum of Hg($^3P_1 \leftarrow\ ^1S_0$) in the $HgAr_2$ cluster.[31] In these systems the ground electronic state of the chromophore atom is spherically symmetric (an $S$ electronic state). The degeneracy of the $P$ excited state of the chromophore is split by interactions with the cluster ligands. However, in the case of clusters containing B, it is the *ground* state which is electronically degenerate. Hence, the absorption spectra of B containing clusters is homologous to that of halogen atoms trapped in noble gas cages, investigated recently by Lawrence and Fajardo.[32]

In the present paper, we report a collaborative experimental and theoretical investigation of the $3s \leftarrow 2p$ transition in the $BAr_2$ cluster. As we describe below, we have observed at high beam source backing pressures several features in the fluorescence excitation spectrum which cannot be assigned to the BAr diatom. Using BAr($X$, $B$) potential energy curves which reproduce our experimental observations on this molecule and an Ar–Ar interaction potential, we have used a pairwise additive model to simulate the structure and electronic absorption of the $BAr_2$ cluster. Our approach parallels closely that of Cheng and Whaley,[30] in our use of variational[33-35] and diffusion[36, 37] Monte-Carlo methods. The simulated spectrum reproduces nearly quantitatively the strongest feature in our experimental spectrum not assignable to BAr. Because of the barrier in the BAr($B^2\Sigma^+$) potential energy curve, the $3s \leftarrow 2p$ transition in the BAr$_2$ is expected to have an asymmetric profile, as is found experimentally.

II. EXPERIMENTAL

The apparatus with which the experimental spectra were obtained has been described in detail previously.[3, 5, 9, 10] A supersonic beam containing B atoms and its weakly bound complexes with one or more Ar atoms was prepared in a pulsed free jet expansion (0.2 mm diam orifice) of B$_2$H$_6$/Ar/He mixtures through 193 nm photolysis of diborane at the nozzle orifice. Fluorescence excitation spectra were recorded 1.2 cm downstream of the nozzle with the frequency-doubled output (bandwidth 0.4 cm$^{-1}$) of a dye laser (Lambda Physik LPD3002E) in the wavelength region near the B atomic $2s^23s\ ^2S \leftarrow 2s^22p\ ^2P$ transition at 249.8 nm. Typical uv pulse energies were 10 $\mu$J in a 0.2 cm diam beam. Fluorescence depletion experiments were carried out by overlapping this tunable laser beam with a second tunable laser beams, as described previously.[10]

The laser-induced fluorescence signal passed through a 1/4 m monochromator and was detected with a photomultiplier (EMI 9813QB), whose output was directed to a gated integrator and thence to a laboratory computer. The monochromator helped to discriminate the fluorescence signal from background light induced by the photolysis laser. In addition, the gain of the photomultiplier was switched off during the excimer laser pulse by zeroing the voltage difference between the photocathode and first dynode. A portion of the fundamental output of the dye laser was directed through a solid-fused silica etalon (free spectral range 1.18 cm$^{-1}$) in order to provide wavenumber markers. The B atomic $2s^23s\ ^2S - 2s^22p\ ^2P_{1/2,3/2}$ transitions[15] were employed for absolute calibration.

We presented in a previous publication[3] a laser fluorescence excitation spectrum of the BAr $B^2\Sigma^+ - X^2\Pi_{1/2}$ transition. We present in Fig. 1 a similar spectrum taken with a high Ar source backing pressure (9.1 atm). Along with the strong B atomic lines, visible in the spectrum are rotationally resolved BAr (v′,0) $B - X$ bands [v′ = 6 and 7], as well as unresolved transitions to higher vibrational levels [v′ = 8 and 9] which lie near or slightly above the barrier in the BAr($B^2\Sigma^+$) potential energy curve.

Also present in the spectrum of Fig. 1 is an additional feature to the blue of the BAr transitions. The intensity of this feature (labeled "BAr$_2$"), relative to the intensities of the BAr bands, was found to be significantly reduced in spectra taken with lower source Ar backing pressures or Ar mole fractions.

We also investigated the molecular carrier of this feature in fluorescence depletion[10, 38, 39] (FD) experiments. This technique is a folded variant of optical-optical double resonance spectroscopy, in which two lasers access the same lower level. This allows the observation of transitions by monitoring the effect of the so-called depletion laser on the fluorescence induced by the probe laser. FD spectroscopy has been employed to detect transitions to non-fluorescing excited states[10, 39, 40] and can also be utilized to detect transitions in specific molecular species.[10, 11, 40] We verified that the feature labeled "BAr$_2$" in Fig. 1 was not due to BAr through FD experiments. Specifically, we scanned the spectral region shown in Fig. 1 with the probe laser set to excite the BAr $D – X$ (0,0) band. [The BAr($D^2\Pi$) state correlates with the excited valence B($2s2p^2$ $^2D$) + Ar asymptote, and the analysis of the $D – X$ transition will be presented in forthcoming publications.[11, 40]] Depletion was observed only on transitions previously identified as belonging to BAr,[3] and not on the "BAr$_2$" feature.

Thus, the molecular carrier of this spectral transition is not the diatomic BAr complex, but a higher BAr$_n$ cluster. In Sec. V, we present a simulated spectrum for the $3s \leftarrow 2p$ transition in BAr$_2$. This simulation reproduces very well the shape and wavenumber range of this experimentally observed feature. In additional FD experiments, we observed the absorption spectrum of the BAr$_2$ complex in the region of the B atomic $2s2p^2$ $^2D \leftarrow 2s^22p$ $^2P$ transition. The interpretation of this spectrum and comparison with a theoretical simulation will be presented in a future publication.

III. BAr COMPLEX

Since our description of the ternary BAr$_2$ complex will be based on a pairwise additive model, we first have refined somewhat our description of the ground ($X^2\Pi$) state and the first bound excited state ($B^2\Sigma^+$) of the binary BAr complex.[3] Our earlier theoretical estimate[3] of the dissociation energy for the $X^2\Pi$ state ($D_0'' = 75.1$ cm$^{-1}$) is ~ 33 % less than a recent experimental estimate by Yang and Dagdigian[40] ($D_0'' = 102.4 \pm 0.3$ cm$^{-1}$). Since the BAr well results from dispersion interactions, the discrepancy between the theoretical and experimental estimates of the well depth is a manifestation of the incomplete recovery of the electronic correlation energy in the theoretical calculations, due to the finite size of the one-electron basis used, and the exclusion of higher order (triple, quadruple, ...) excitations out of the reference space.

In earlier investigations of the BNe and $C_2H_2Ar$ complexes,[5, 41] we suggested that this shortcoming of the theoretical calculations could be easily corrected by a simple scaling of the correlation energy. This had already been proposed by Brown and Truhlar[42] and Sölter et al.[43] in calculations on chemically bonded systems. Here we define the correlation energy as the difference between the multi-reference, averaged coupled-pair (MR-ACPF) energy and the energy expectation value of the CASSCF reference function from our earlier *ab initio* calculations on BAr($X^2\Pi$).[3] In terms of this correlation energy, the scaled BAr($X^2\Pi$) potential curve is defined as

$$E_s(R) = E_{CASSCF}(R) + s\, E_{CORR}(R) \ . \tag{1}$$

A proper treatment of the vibration-rotation levels of the B(2p)Ar complex proceeds through a full close-coupled expansion of the wavefunction[44]

$$\Psi^{JM}(\mathbf{R}) = \sum_{jL} C_{jL}^{JM}(R)(jm_j\, LM_L\,|\,JM)|LM_L>|l\,s\,j\,m> \ . \tag{2}$$

Here $l$, $s$, $j$ are the electronic orbital, spin, and total angular momentum of the B atom; $L$ is the orbital angular momentum for the rotation of the B–Ar pair, and $J$ is the total angular momentum. The space-fixed projections of $j$, $L$, and $J$ are denoted $m_j$, $M_L$, and $M$, respectively. The expansion coefficients $C_{jL}^{J}(R)$ satisfy the close-coupled equations

$$\left[\frac{-\hbar^2}{2\mu}\left(\frac{d^2}{dR^2}\mathbf{1} - \frac{\mathbf{L}^2}{R^2}\right) + \tilde{\mathbf{V}}(R) - E\mathbf{1}\right]\mathbf{C}(R) = 0 \ , \tag{3}$$

where $\mathbf{1}$ is the unit matrix, $\mathbf{L}^2$ is the diagonal matrix of the orbital angular momentum, with elements $L(L+1)$, and $\tilde{\mathbf{V}}(R)$ is the full matrix of the electrostatic interaction potential plus the spin-orbit operator in this $\{\,j\,L\,J\,M\,\}$ basis.

For each value of the total angular momentum $J$ ($J > 1/2$), there are 6 allowed $|\,j\,L\,J\,M>$ states, which separate into two non-interacting sets of 3 states, one set for each allowed value of the total parity.[44] For the lowest rotational level of the B(2p)Ar complex ($J = 1/2$) the problem decouples further, into two sets of 2 states, and one state which is uncoupled.[44] The two sets of $2 \times 2$ potential matrices are identical and are given by Eq. (A5) in Appendix A. The individual elements of this matrix involve the potential energy curves for the $X^2\Pi$ and $A^2\Sigma^+$ states of BAr,[3] which we denote $V_\Pi(R)$ and $V_\Sigma(R)$, respectively, as well as the spin-orbit constant of the B atom in the ground $(2s^2 2p\ ^2P)$ electronic state [$a = 10.169$ cm$^{-1}$ (Ref. 15)]. The vibronic (vibrational-rotational-electronic) states of BAr correspond to the bound-state solutions of Eq. (3).

The description can be simplified by the adiabatic approximation, in which the vibronic states correspond to solutions of the following uncoupled equation, which corresponds to motion governed by a single potential energy curve,

$$\left[ \frac{-\hbar^2}{2\mu}\left(\frac{d^2}{dR^2} - \frac{L(L+1)}{R^2}\right) + \tilde{V}_{ad}(R) - E \right] C_L(R) = 0 , \qquad (4)$$

where $\tilde{V}_{ad}(R)$ is the lowest eigenvalue of the $\tilde{\mathbf{V}}(R)$ matrix. The BAr($X^2\Pi$) rotational constant is given by

$$B = \frac{\hbar^2}{2\mu} \int |C_L|^2 R^{-2}\, dR . \qquad (5)$$

With a choice of scaling factor $s = 1.19$, the calculated dissociation energy and rotational constants for BAr($X^2\Pi$), 106.1 cm$^{-1}$ and 0.149 cm$^{-1}$, respectively, agree extremely well with the experimental estimates:[3, 40] $102.4 \pm 0.3$ cm$^{-1}$ and 0.150 cm$^{-1}$. The exact energy, given by solution of Eq. (3), differs from the adiabatic estimate by $< 0.02$ cm$^{-1}$, which indicates that nonadiabatic effects are very small indeed. Lowering the scaling factor slightly has the effect of decreasing the error in the calculated dissociation energy, but increasing the error in the calculated value of the rotational constant. An identical scaling factor ($s = 1.19$) was used to define the potential energy curve for the mainly repulsive BAr($A^2\Sigma^+$) state, which also correlates with the ground atomic asymptote B($2s^2 2p\ ^2P$) + Ar. Table I gives values for $V_\Pi(R)$ and $V_\Sigma(R)$, for a scaling factor of 1.19.

The BAr($B^2\Sigma^+$) state correlates with the B($2s^2 3s\ ^2S$) + Ar asymptote. To define the potential energy curve for this state, which we designate $V_{3s}(R)$, we use the experimental RKR potential[3] inside the region defined by the bound vibrational levels of this state (v' ≤ 6). With our recent experimental estimate of the BAr($X^2\Pi$) dissociation energy,[40] our previously measured transition wavenumbers for $B - X$ (v',0) bands,[3] and the B atomic transition wavenumber [$h\nu$ = 40039.7 cm$^{-1}$ (Ref. 15)], the RKR potential could be fixed relative to the energy of the B($2s^2 3s\ ^2S$) + Ar asymptote. The height, width, and general shape of the barrier were then adjusted so that the calculated positions and widths of the metastable states trapped by this barrier (v' = 7 – 9) agreed well with the experimental estimates. Table II gives values for the resulting $V_{3s}(R)$ potential energy curve. Figure 2 displays the potential energy curves of the $X^2\Pi_{1/2,\ 3/2}$, $A^2\Sigma^+$, and $B^2\Sigma^+$ states of BAr. The height of the barrier in the $B$ state potential energy curve is 108.0 cm$^{-1}$, at $R$ = 7.178 bohr. The dissociation energy for $V_{3s}(R)$ is $D_0$ = 1000.5 cm$^{-1}$.

The calculated positions and rotational constants for the bound vibrational states of the $^{10}$BAr($B^2\Sigma^+$) and $^{11}$BAr($B^2\Sigma^+$) isotopomers, and the positions and widths of the predissociating states (determined as described in Ref. 3) are compared in Table III with the experimental estimates[3, 40] of these quantities. As can be seen, the agreement is excellent. The position of the BAr($B^2\Sigma^+$) bound and predissociating states are also shown in Fig. 2.

IV. BAr$_2$ COMPLEX: STRUCTURE

We construct the potential for the ternary BAr$_2$ complex using a pairwise additive model. As will be discussed below, we shall follow an adiabatic approach, involving a sum of pairwise potential matrices, to account for the orientational property of the singly filled 2$p$ orbital of the B atom. Our treatment is equivalent to that introduced by Balling and Wright[45] to simulate the absorption spectra of alkali atoms in noble gas matrices. This treatment of the interaction between an atom in a $P$ electronic state with an ensemble of spherical perturbers has been used subsequently by Boatz and Fajardo,[29] Lawrence and Apkarian,[32] Zúñiga et al.,[31] and Cheng and Whaley.[30] We assume, without loss of generality, that the B atom is fixed at the origin of the coordinate system. As discussed in the preceding section and in Appendix A, the interaction of an Ar atom with the B atom can be expressed by a 6 × 6 matrix using the uncoupled (or coupled) spin-orbit states of the B($2s^2 2p\ ^2P$) atom as a basis. If the Ar atom is located at a distance $R$ along the $z$-axis, this matrix, designated **V**($R$) in Appendix A, can be additionally block diagonalized.

If the Ar atom is arbitrarily positioned in space, at polar coordinates $(R, \theta, \phi)$, the matrix describing its electrostatic interaction with the B atom is given by

$$\mathbf{V}(R, \theta, \phi) \equiv \mathbf{D}(\phi, \theta, 0) \, \mathbf{V}(R) \, \mathbf{D}^T(\phi, \theta, 0) \ , \tag{6}$$

where $\mathbf{D}(\phi, \theta, 0)$ is the matrix, in the basis of the spin-orbit states of the B atom, of the rotation specified by the Euler angles $\{\phi, \theta, 0\}$. The matrix $\mathbf{V}(R)$ is the matrix of the interaction potential *not including* the spin-orbit interaction. As described in more detail in Appendix A, $\mathbf{V}(R, \theta, \phi)$ is the $6 \times 6$ matrix, in an uncoupled Cartesian basis, for the electrostatic interaction of an Ar atom at polar coordinates $(R, \theta, \phi)$ with a B atom sitting at the origin with 2p orbital oriented along the *x*, *y*, or *z* axes.

In the adiabatic approximation, which we have seen was of high accuracy for the binary BAr complex, the $BAr_2$ complex is characterized by a single potential energy surface, $V_{ad}(R_1, \theta_1, \phi_1, R_2, \theta_2, \phi_2, R_{12})$, where the subscripts 1 and 2 designate the two Ar atoms and $R_{12}$ is the distance between these two Ar atoms. Of course, only the $R_i$, $\theta_i$, and $\phi_i$ are independent variables. The potential energy surface $V_{ad}$ is the lowest root of the $6 \times 6$ matrix

$$\mathbf{V}(R_1, \theta_1, \phi_1) + \mathbf{V}(R_2, \theta_2, \phi_2) + \mathbf{a} + V_{Ar_2}(R_{12}) \, \mathbf{1} \ , \tag{7}$$

which describes the electrostatic and spin-orbit interactions in the $BAr_2$ system. In Eq. (7), **a** is the matrix of the spin-orbit operator in the same uncoupled Cartesian basis (see Appendix A). Also, $V_{Ar_2}(R_{12})$ is the scalar interaction potential between two Ar atoms separated by a distance $R_{12}$. We use the $Ar_2$ potential of Aziz and Chen,[46] with $R_e = 7.1053$ bohr, $D_e = 97.81$ cm$^{-1}$, and $D_0 = 83.1$ cm$^{-1}$.

Since the potential energy surface is taken as pairwise additive, the minimum energy of the $BAr_2$ cluster is attained in an isosceles triangular configuration, with $R_1 = R_2$ and $R_{12}$ equal to the equilibrium internuclear separation in the $Ar_2$ cluster. Since three atoms define a plane, both Ar atoms can be situated in a plane perpendicular to the B 2p orbital, which is the optimal geometry. Thus we anticipate, and find, that the BAr distance in the $BAr_2$ cluster is identical to the equilibrium internuclear separation in the binary BAr cluster, 6.6889 bohr. The energy of the $BAr_2$ cluster at its equilibrium geometry is –356.9 cm$^{-1}$.

Within this adiabatic approximation, the nuclear motion of the $BAr_2$ complex is the solution to the Schrödinger equation

$$\left[ -\frac{\hbar^2}{2} \sum_{i=1}^{3} \frac{1}{m_i} \nabla_i^2 + V_{ad}(R_1, \theta_1, \phi_1, R_2, \theta_2, \phi_2, R_{12}) - E \right] \Psi = 0 \; , \qquad (8)$$

where the sum runs over the three nuclei. The eigenenergy $E$ of Eq. (8) equals $-D_0$, where $D_0$ is the dissociation energy of the $BAr_2$ complex, corrected for zero-point motion. To determine this energy we use first a variational Monte-Carlo (VMC)[33-35] method, followed by a diffusion Monte-Carlo (DMC)[36, 37] method.

**A. Variational Monte-Carlo method**

Defining

$$H(\boldsymbol{q}) = -\frac{\hbar^2}{2} \sum_{i=1}^{3} \frac{1}{m_i} \nabla_i^2 + V_{ad}(R_1, \theta_1, \phi_1, R_2, \theta_2, \phi_2, R_{12}) \; , \qquad (9)$$

where $\boldsymbol{q}$ represents the coordinates of all three particles, we write the variational approximation to the energy $E$ in the following form

$$E \leq E_{var} \equiv \int \Phi(\boldsymbol{q})^* \Phi(\boldsymbol{q}) E_{loc}(\boldsymbol{q}) d\boldsymbol{q} \Big/ \int \Phi(\boldsymbol{q})^* \Phi(\boldsymbol{q}) \, d\boldsymbol{q}, \qquad (10)$$

where $\Phi(\boldsymbol{q})$ is an approximate (variational) wavefunction. The local energy $E_{loc}$ in Eq. (10) is defined by[47]

$$E_{loc}(\boldsymbol{q}) \equiv H(\boldsymbol{q}) \, \Phi(\boldsymbol{q}) / \Phi(\boldsymbol{q}) \; . \qquad (11)$$

In the VMC method, both integrals in Eq. (10) are evaluated numerically,

$$E_{var} \cong \sum_{i=1}^{N} \Phi(q_i)^* \Phi(q_i) E_{loc}(q_i) \Big/ \sum_{i=1}^{N} \Phi(q_i)^* \Phi(q_i) \; , \qquad (12)$$

where the points $\{q_i\}$ are chosen to be appropriate to the probability distribution

$$P(q) \equiv \Phi(q)^* \Phi(q) \ . \tag{13}$$

To choose these points we use the standard Metropolis algorithm, as follows. An ensemble of $N$ sets of Cartesian coordinates ($q_i$) for all three particles is chosen using a uniform random number generator over a range $\pm \, dq$ centered on the equilibrium position of each particle. The local energy is then evaluated. In this procedure, the standard three-point finite-difference approximation[48] was used to evaluate the Laplacian. Each set of Cartesian coordinates is then changed by addition of a random increment chosen from a uniform random distribution centered at zero and spanning the range $\pm \, \delta$. If the probability $P(q)$ is larger for the randomly altered set of coordinates, the old set is replaced by the randomly altered set. If the probability is smaller for the randomly altered set of coordinates, then the randomly altered set is retained only if the ratio of the probability at the new set of coordinates divided by the probability at the old set of coordinates is greater than a uniform random number chosen on the interval {0,1}. The local energy is then re-evaluated and the whole procedure iterated.

The numerical (exact) wavefunctions for the lowest vibrational states of the BAr($X$) and Ar$_2$ complexes can be approximated extremely well ($S > 0.9999$) by Morse functions. The Morse parameters which define the best fit to the numerical (exact) wavefunctions are given in Table IV. Since a Morse function does so well in approximating the vibrational wavefunctions of the embedded binary complexes, a good choice for a variational approximation to the wavefunction of the vibrational ground state of the ternary B(2$p$)Ar$_2$ complex will be a product of three Morse functions, namely

$$\begin{aligned}\Phi_g(q) = \ & \Phi_M(R_{12}; R_{e1}, D_{e1}, \beta_1, \mu_1) \, \Phi_M(R_{13}; R_{e1}, D_{e1}, \beta_1, \mu_1) \\ & \times \Phi_M(R_{23}; R_{e2}, D_{e2}, \beta_2, \mu_2) \ ,\end{aligned} \tag{14}$$

where the subscript indices "1" and "2" on the Morse parameters refer, respectively to the BAr and Ar$_2$ systems, and the three coordinates $R_{12}$, $R_{13}$, and $R_{23}$ refer, respectively, to the BAr$_1$, BAr$_2$, and Ar$_1$Ar$_2$ separations.

In Eq. (14) $\Phi_M(R; R_e, D_e, \beta, \mu)$ designates the vibrational ground state wavefunction for a Morse potential

$$V(R) = D_e \{ \exp[-2\beta(R - R_e)] - 2\exp[-\beta(R - R_e)] \}, \qquad (15)$$

namely

$$\Phi_M(R; R_e, D_e, \beta, \mu) = \left[\frac{2\beta(K - 1/2)}{\Gamma(2K)}\right]^{1/2} \exp(-x/2) \, x^{K-1/2}, \qquad (16)$$

where

$$K = (2 D_e \mu)^{1/2} / \beta, \qquad (17a)$$

and

$$x = 2 K \exp[-\beta(R - R_e)]. \qquad (17b)$$

For $BAr_2$ we choose, in Eqs. (16) and (17), the Morse parameters for the embedded binary complexes, from Table IV. The three parameters which control the VMC integration, $N$, $dq$, and $\delta$, are adjusted empirically so that *(a)* this convergence is rapid, *(b)* the residual statistical fluctuation is within an acceptable limit of uncertainty in the calculated energy, and *(c)* the rate of acceptance of the randomly altered coordinates is ~ 50 %. We have chosen $dq = 4$ and $\delta = 3/8$ (both in bohr). Figure 3 illustrates the convergence of the variational energy [Eq. (12)] as the number of iterations increases, using both 100 and 500 integration points $N$, and $dq = 4$ and 5. We observe in Fig. 3 that the magnitude of the statistical fluctuations in the energy, once convergence has been reached, vary as ~ $N^{-1/2}$, as expected. As an improved estimate of the energy, we average the calculated energy of five independent runs with $N = 500$ and $dq = 4$, obtaining $E = -295.12 \pm 0.21$ cm$^{-1}$.

In an attempt to improve the variational energy, we implemented a Monte-Carlo based variation[49] of the 6 parameters which define the trial wavefunction for the $BAr_2$ cluster (Table IV). The $n^{th}$ parameter, selected randomly, was allowed to vary by an amount, also selected randomly from a uniform distribution in the range ± 10% of the original value of the parameter. The variational energy for the new set of parameters is evaluated using the combined set of integration points corresponding to the last 10 Metropolis sampling iterations with the original set of parameters. Since the integration points are not readjusted (resampled) for the new set of parameters, this evaluation of the variational energy is called "reweighting."[34]

Repeated applications of this Monte-Carlo variation of the parameters used to define the trial wavefunction never yielded an energy which was lower, to within statistical error, than the calculated energy for the product of the Morse functions for the embedded binary complexes, defined by the parameters shown in Table IV. This energy ($-295.12 \pm 0.21$ cm$^{-1}$) is the VMC approximation to the energy of the vibrational ground state of the B($2p$)Ar$_2$ cluster.

The calculated dissociation energies of the binary B($2p$)Ar and Ar$_2$ clusters are (see Sec. III) $D_0 = 106.1$ cm$^{-1}$ and 83.1 cm$^{-1}$, respectively. Thus, as we might have anticipated, the calculated VMC energy of the vibrational ground state of B($2p$)Ar$_2$ is, to within statistical error, equal to the sum of the dissociation energies of the three bonds.

## B. Diffusion Monte-Carlo method

The Diffusion Monte-Carlo (DMC) method[36, 37] uses the fact that the time-dependent Schrödinger equation may be converted into a diffusion equation by making a transformation to imaginary time. The oscillatory solutions of the time-dependent Schrödinger equation are thereby mapped onto a set of relaxation processes of the form

$$\Phi(\boldsymbol{q}, \tau) = \sum_n c_n \phi_n \exp[-(E_n - E_r)\tau] \ , \qquad (18)$$

where $\phi_n$ and $E_n$ are the eigenfunctions and eigenvalues, respectively, of the Hamiltonian in question (which is taken to be time independent) and the zero of energy has been redefined as $E_r$. If this "reference energy" is the ground state energy, then all the components of Eq. (18) will decay to zero except for the ground state eigenfunction.

The DMC method is implemented using an unguided random walk. An ensemble of $N_0$ sets of Cartesian coordinates ($q_i$) for all three particles is chosen. At each time step, all members of this ensemble, designated "replicas", are displaced in space, with the displacement of the $k^{th}$ Cartesian coordinate of the $i^{th}$ replica chosen randomly from a normal distribution with standard deviation $\sigma_k$ proportional to the square root of the product of the (imaginary) time step and the diffusion coefficient[36]

$$\sigma_k = (2 D_k \Delta\tau)^{1/2} \ , \qquad (19)$$

where the latter is given by

$$D_k = \hbar^2 / 2m_k \ . \tag{20}$$

Here the subscript refers to the $k$th Cartesian coordinate and $m_k$ is the mass of the particle associated with this coordinate.

Initially equal weights are assigned to each replica. In the "continuous weighting" application of the DMC method,[37] after every time step $\Delta\tau$, the weight of every replica is adjusted following the algorithm

$$w_i(\tau) = P_i \, w_i(\tau - \Delta\tau) \ . \tag{21}$$

where

$$P_i = \exp\{[E_r - V(\mathbf{q}_i)]\Delta\tau\} \ . \tag{22}$$

If the weight of any replica drops below a certain value (here taken to be $1/N$), then that replica is eliminated from the ensemble. A replacement is created by duplicating the replica with the highest weight, halving the weights of both the duplicate and the original, and adding the weight of the eliminated replica to the duplicate. In this way the total number of replicas remains constant.

Following the readjustment of the weights, the reference energy, $E_r$, is updated. We use the simple algorithm proposed by Anderson[36]

$$E_r(\tau) = \langle V(\tau) \rangle - a[N(\tau) - N_0]/N_0 \ , \tag{23}$$

where $N_0$ is the sum of the initial weights of the replicas (equal to the ensemble size if the weights of each of the replicas is initially set equal to 1), $N(\tau)$ is the sum of the replica weights at time $\tau$ and $\langle V(\tau) \rangle$ is the mean potential energy given by

$$\langle V(\tau) \rangle = \sum_i w_i \, V(\mathbf{q}_i) / N(\tau) \ . \tag{24}$$

The value of $a$ in Eq. (23) depends on the size of the time step:

$$a = min(\Delta\tau, \Delta\tau^{-1}) , \qquad (25)$$

with $\Delta\tau$ measured in atomic units. Finally, the kinetic energy associated with the translational motion of the cluster is subtracted at the end of each displacement by repositioning the center-of-mass of each cluster (replica) at the origin.

The DMC method provides two estimates for the ground state energy, namely the reference energy, $E_r$, and the mean potential energy, $\langle V(\tau) \rangle$. For the calculations reported here, the ground state energy was taken to be the time average of $\langle V(\tau) \rangle$ over the last 10% of the total propagation time. To construct a histogram approximation to the ground state wavefunction, the three interatomic distances and the replica weights were summed over the last 10% of the total propagation time.

Both the ground state energy and probability density can be calculated more efficiently by importance sampling.[37, 50-52] We define a new function $f = \Phi\Phi_t$, where $\Phi_t$ is a trial wavefunction. Substitution for $\Phi$ in the time-dependent Schrödinger equation (written in imaginary time) results in the equation for $f$

$$\frac{df}{d\tau} = \sum_k \frac{\hbar^2}{2m_k} \nabla_k^2 f - \sum_k \nabla_k \cdot (f \nabla_k ln \Phi_t) - \left(\frac{H\Phi_t}{\Phi_t} - E_r\right)f . \qquad (26)$$

The first term on the right hand side of Eq. (26) represents the diffusion of the replicas in configuration space and may be treated as described earlier in this section. Here, however, the replicas correspond to the discrete representation of $f$ rather than the ground state wavefunction. The second term corresponds to a drift of the particles in a fluid of velocity $\nabla_k ln \Phi_t$.[50] To accommodate this term, after the random diffusion displacement, the replicas are subjected to an additional displacement, given by the product of the drift velocity and the time step, $(\nabla_k ln \Phi_t)\Delta\tau$.

After every time step the weight of every replica is still readjusted according to Eq. (21) but with the transition probabilities now defined by

$$P_i = exp\{[E_r - H\Phi_t/\Phi_t]\Delta\tau\} \quad , \tag{27}$$

where the potential in Eq. (22) is replaced by the local energy of the trial function. If this latter is identical to the ground state wavefunction, then the last term on the right hand side of Eq. (26) will vanish when $E_r = E_0$. Further, the reference energy for each replica is defined initially as the energy of the cluster and $E_r$ is still adjusted according to Eq. (23) which makes use of the mean potential energy, $\langle V(\tau) \rangle$, .

In our implementation of importance sampling, the trial wavefunction was taken to be the product of Morse functions used in the VMC calculations (discussed in the preceding sub-section and defined by the parameters in Table IV). We determine the DMC approximation to the ground state energy as before, averaging the mean potential energy over the last 10% of the time steps. However, it is the function $f$ which is now represented by the replica weights. Hence, we construct $\Phi$ from the ratio $\Phi = f/\Phi_t$.[14] Then, exactly as in the determination of $\Phi$ without the use of importance sampling, the importance sampled approximation $\Phi$ is averaged over the last 10% of the total propagation time.

As discussed above, the ground-state vibrational wavefunctions for both the B(2p)Ar and B(2p)Ar$_2$ complexes may be accurately approximated by Morse functions. Thus, for the DMC calculations reported here, the importance sampling technique was implemented taking the wavefunction given in Eq. (14) as the trial wavefunction. Each replica was initially defined as having the equilibrium geometry, with a small random displacement from a normal distribution with the standard deviation defined in Eq. (19). An ensemble of 2000 replicas was propagated for a total time of 40000 atomic time units, using a time step of 0.25 a.u.

The lower panel of Fig. 3 compares the calculated values of the reference energy as a function of imaginary time, for a representative DMC run, with and without importance sampling. As can be seen, the use of importance sampling dramatically improves the convergence. The final energy from the DMC + importance sampling calculation is $-295.9 \pm 0.1$ cm$^{-1}$. This is ~ 0.8 cm$^{-1}$ lower than the VMC value – a slight improvement.

## V. BAr$_2$ COMPLEX: SPECTRUM

As can be seen from Fig. 2, electronic excitation of the B atom in the BAr$_2$ complex, at its equilibrium geometry, will lead to formation of the B(3s)Ar$_2$ complex with the B(3s) atom located just slightly inside the top of the barrier in the B(3s)Ar potential energy curve. The excited state, so produced, will, however, not be stable with respect to elimination of an Ar atom since its energy is greater than that of the B(3s)Ar + Ar asymptote. Because the B(3s)Ar well depth is so deep ($D_0 = 1000.5$ cm$^{-1}$), decreasing the length of one of the BAr bonds will provide more than enough energy to stretch the second BAr bond over the barrier and also break the Ar–Ar bond. Thus electronic excitation of the BAr$_2$ complex will correspond to a free $\leftarrow$ bound photodissociation.

To simulate the spectrum, we use the semiclassical expression,[30, 53]

$$I(\omega) \sim \int \left| \Phi_g(\mathbf{q}) \right|^2 \delta\left[ V_{3s}(\mathbf{q}) - V_{2p}(\mathbf{q}) - \hbar\omega \right] d\mathbf{q} \quad , \tag{28}$$

where $\Phi_g$ and $V_{3s}(\mathbf{q})$ are the wavefunction and potential energy, respectively, of the vibrational ground state of the B(2p)Ar$_2$ complex and $V_{3s}(\mathbf{q})$ is the potential energy of the B(3s)Ar$_2$ complex. Since B(3s) is electronically non-degenerate, within a pairwise additive approximation we represent $V_{3s}(\mathbf{q})$ – the PES for the B(3s)Ar$_2$ complex – as a sum of two B(3s)Ar potentials defined in Sec. III and the Ar$_2$ potential of Aziz and co-workers.[46] Since we are using a pairwise additive description of both the ground and excited states and since the Ar$_2$ interaction appears as a scalar term in the description of the ground state [see Eq. (7)], the Ar–Ar potential energy cancels inside the delta function in Eq. (28).

In Eq. (28), we have ignored the dependence of the 3s $\leftarrow$ 2p electronic transition moment on the B–Ar separations. We previously found[4] that the BAr $B - X$ and $B - A$ transition moments remained equal to their asymptotic, atomic values and were independent of the internuclear separation $R$, except for very small $R$.

Consistent with our Monte-Carlo treatment, we approximate the integral in Eq. (18) as

$$I(\omega) \sim \sum_{i=1}^{N} \left| \Phi_g(q_i) \right|^2 \delta\left[ V_{3s}(q_i) - V_{2p}(q_i) - \hbar\omega \right] \quad . \tag{29}$$

The Kronecker delta in Eq. (29) is evaluated by a boxcar histogram: The spectral region under investigation is divided into equally spaced bins, of width $\delta\omega$. For each integration point, the weight $\left|\Phi_g(q_i)\right|^2$ is assigned to the particular bin in which the quantity $V_{3s}(q_i) - V_{2p}(q_i)$ lies. To reduce statistical error, the integration points in Eq. (29) include the results of five independent runs with $N = 500$. The "noise" in the predicted spectrum varies inversely with the bin width. A bin width of 2 cm$^{-1}$ resulted in structure in the calculated spectra comparable with that seen in the experiment.

The classical equivalent to the absorption profile given by Eq. (28), in which $\left|\Phi_g(\boldsymbol{q})\right|^2$ is replaced by the statistical mechanical probability distribution, was used by Boatz and Fajardo,[29] and Lawrence and Apkarian.[32] Equation (28) has been discussed in detail by Cheng and Whaley[30] and compared with earlier equivalent expressions[54] based on the semiclassical "reflection" approximation. Cheng and Whaley[30] suggest that Eq. (28) can be improved by explicitly correcting for the kinetic energy of the B chromophore in the excited state, which gives[30]

$$I(\omega) \sim \int \left|\Phi_g(\boldsymbol{q})\right|^2 \delta\left[V_{3s}(\boldsymbol{q}) - T_B(\boldsymbol{q}) - V_{2p}(\boldsymbol{q}) - \hbar\omega\right] d\boldsymbol{q} \quad . \tag{30}$$

We approximate this kinetic energy term as

$$T_B(\boldsymbol{q}) = -\frac{\hbar^2}{2m_B}\left[\sum_i \nabla_i^2 \Phi_g(\boldsymbol{q})\right] \Big/ \Phi_g(\boldsymbol{q}) \quad , \tag{31}$$

where the sum runs over the three Cartesian coordinates of the B atom. Equation (31) is just the component of the local energy in the ground state associated with the kinetic energy of the B atom.

Figure 5 presents the calculated B($3s \leftarrow 2p$)Ar$_2$ spectra. For the VMC simulation, the integration points $\{q_i\}$ in Eq. (29) were determined by combining the final integration points from four separate VMC calculations with $N = 500$ and $dq = 4$ (see Sec. III A), for a combined total of 2000 integration points. For the importance sampled DMC simulation 2000 replicas were propagated for 40 000 steps in imaginary time. To simulate the spectra, while reducing any effect of correlation between individual time step, we accumulated the coordinates of each replica and their weights at steps 37 000, 38 000, 39 000, and 40 000; for a total of 8000 points. The kinetic energy correction for the DMC spectral simulation [Eq. (31)] was evaluated using the Morse product function as $\Phi_g(\mathbf{q})$.

Figure 5 also presents the experimental spectra associated with the BAr$_2$ complex. As discussed in the experimental section, at energies below ~ 290 cm$^{-1}$ [relative to the B(3s)+Ar+Ar asymptote], the feature associated with BAr$_2$ is obscured by the wing of the BAr ($B^2\Sigma^+$, v' = 9) ← ($X^2\Pi_{1/2}$, v" = 0) transition. This transition occurs at

$$\hbar\omega = 128 \text{ (Table III)} + D_0[\text{BAr}(X)] = 230 \text{ cm}^{-1}, \qquad (32)$$

relative to the B $3s\ ^2S \leftarrow 2p\ ^2P_{1/2}$ transition wavenumber and is substantially broadened by predissociation.

The agreement with experiment is excellent. We observe that the kinetic energy correction, as defined by Eqs. (30) and (31), shifts the peak of the calculated spectrum slightly to lower energy. The asymmetry in the line shape, which is clearly present in both the DMC and VMC simulations, is a consequence of the shape of the potential energy difference, $V_{3s}(q_i) - V_{2p}(q_i)$, which appears in Eq. (28), in the region sampled by the ground state vibrational wavefunction. This is illustrated by Fig. 6, which displays contours of both $V_{3s}(q_i) - V_{2p}(q_i)$ and $|\Phi_g(\mathbf{q})|^2$. In fact, the intensity at frequency $\omega$ is given by the surface integral

$$I(\omega) = \iint_A |\Phi_g(\mathbf{q})|^2 d\mathbf{q} \ , \qquad (33)$$

where the surface of integration is defined by $V_{3s}(q_i) - V_{2p}(q_i) = \hbar(\omega \pm \delta\omega)$. The asymmetry in the lineshape, and the sharp cutoff, reflects the occurrence of a maximum in the potential energy difference at ~ 450 cm$^{-1}$, which nearly coincides with the maximum in the square of the ground state vibrational wavefunction.

V. DISCUSSION

As can be seen in Fig. 5, the VMC and importance-sampled DMC calculations predict an excitation spectrum for the $3s \leftarrow 2p$ transition of the BAr$_2$ complex that agrees well, both in location, width, and asymmetry, with the experimentally observed spectrum. The agreement is somewhat better with the VMC calculation, which indicates that this latter approach likely provides a better description of the ground state wavefunction and/or a better prescription for the evaluation of the semi-classical integral for the spectral intensity [ Eq. (29) ].  The good agreement between the predicted and observed spectra is further confirmation that this feature does indeed correspond to the B($3s \leftarrow 2p$) excitation within the BAr$_2$ complex.

Introduction of the kinetic energy correction of Eq. (30) leads to a closer match between the VMC simulation and the experimental spectrum.  The agreement is, we feel, excellent.  The residual differences could reflect remaining errors in the semiclassical approximation to the absorption lineshape.

Another origin for the difference between the predicted (VMC) and experimental spectra would be inaccuracies in the two-body potentials.  In particular, we see from Figs. 2 and 4 that a vertical B($3s \leftarrow 2p$) transition within the BAr$_2$ complex will probe the B($3s$)Ar potential in the region of the barrier, where it is less well determined spectroscopically.

A third origin of the difference between the predicted (VMC) and experimental spectra is errors resulting from use of the pairwise construction of the matrix of the electronic Hamiltonian and the use of the lowest root of this Hamiltonian to define the adiabatic potential energy surface.  As discussed in Sec. III, we anticipate on the basis of our calculations on B($2p$)Ar, that nonadiabatic effects will be negligibly small for the ternary complex.

Notwithstanding the slight residual difference between the calculated and experimental $3s \leftarrow 2p$ absorption spectrum of the BAr complex, we are extremely pleased by the overall good agreement.  This article shows the degree to which a sophisticated theoretical investigation can enhance the value and understanding of experiment.  Indeed, without comparison of the experimental results with such theoretical calculations, assignment of the observed feature in our experimental spectrum to the BAr$_2$ complex would have been to make convincingly.

Our success in applying Monte-Carlo methods, particularly the VMC method, to the study of complexes involving the open-shell B atom, encourages the study of larger clusters.  Hopefully, these calculations will guide the way for experimental investigation of such clusters and provide an assignment of the $3s \leftarrow 2p$ transition for boron atoms embedded in bulk matrices.

ACKNOWLEDGMENT

This research was supported by the Air Force Office of Scientific Research under Grant Nos. F49620-95-1-0046 and F49620-95-1-0099 (to M.H.A.) and F49620-95-1-0055 (to P.J.D.). Millard Alexander is extremely indebted to Stéphane Guérin and Birgitta Whaley for having stimulated and encouraged his interest in the variational Monte-Carlo method, and to Jerry Boatz, Stuart Rothstein, and Peter Langenfelder for helpful discussions.

APPENDIX A: MATRIX OF THE INTERACTION POTENTIAL

Consider the interaction between an atom in a $^2P$ electronic state and a closed-shell atom which lies along the $z$-axis. The atom may be described in a coupled ($|\, l\, s\, j\, m >$) or uncoupled ($|\, l\, m_l\, s\, m_s >$) basis.[55] In matrix notation the transformation between these two representations can be written

$$\{l\, s\, j\, m\} = \mathbf{C}\, \{l\, m_l\, s\, m_s\}\ ,  \tag{A1}$$

where $\mathbf{C}$ is a $6 \times 6$ matrix of Clebsch-Gordan coefficients.[55] In the coupled basis the spin-orbit operator is diagonal, with matrix elements

$$< l\, s\, j\, m\, |\, H_{so}\, |\, l'\, s'\, j'\, m' > = \tfrac{1}{2} \delta_{j j'} \delta_{mm'}\, a\, [\, j(j+1) - s(s+1) - l(l+1)\, ]\ .  \tag{A2}$$

By contrast, the electrostatic interaction is diagonal in the *uncoupled* basis, with matrix elements given by

$$< l\, m_l\, s\, m_s\, |\, V\, |\, l\, m_l'\, s\, m_s' > = \delta_{m_s m_s'}\, \delta_{m_l m_l'}\, [\, \delta_{m_l,0}\, V_\Sigma(R) + \delta_{|m_l|,1}\, V_\Pi(R)\, ]\ ,  \tag{A3}$$

where $V_\Sigma(R)$ and $V_\Pi(R)$ are the Hund's case (a) B($2s^2 2p$)Ar electrostatic potentials for an orientation in which the Ar atom approaches, respectively, colinearly or perpendicularly to the singly-filled $2p$ orbital.

It will also be convenient to use an uncoupled, Cartesian basis in which the $|l\, m_l\rangle$ states are replaced by $|l\, q_l\rangle$ states, where $q_l = x$, $y$, or $z$ and specifies the orientation of the real (Cartesian) B 2p orbitals. Keeping a notation similar to that used in Sec. III, in this uncoupled, Cartesian basis we designate the full $6\times 6$ matrix of the electrostatic interaction potential plus the spin-orbit operator as $\tilde{\mathbf{V}}(R)$, which is given by

| $q_l$ | $m_s$ | $q_l = x$, $m_s = 1/2$ | $y$, $1/2$ | $z$, $1/2$ | $x$, $-1/2$ | $y$, $-1/2$ | $z$, $-1/2$ |
|---|---|---|---|---|---|---|---|
| $x$ | $1/2$ | $V_\Pi(R)$ | $ia/2$ | $0$ | $0$ | $0$ | $a/2$ |
| $y$ | $1/2$ | | $V_\Pi(R)$ | $0$ | $0$ | $0$ | $ia/2$ |
| $z$ | $1/2$ | | | $V_\Sigma(R)$ | $ia/2$ | $a/2$ | $0$ |
| $x$ | $-1/2$ | | | | $V_\Pi(R)$ | $ia/2$ | $0$ |
| $y$ | $-1/2$ | | | | | $V_\Pi(R)$ | $0$ |
| $z$ | $-1/2$ | | | | | | $V_\Sigma(R)$ |

(A4)

The matrix is Hermitian; for simplicity we have not shown the lower triangle.

In fact, by rearranging the states, this matrix can be factored into two identical $3\times 3$ blocks, the first coupling the states $|x,1/2\rangle$, $|y,1/2\rangle$, and $|z,-1/2\rangle$ and the second coupling the states $|x,-1/2\rangle$, $|y,-1/2\rangle$, and $|z,1/2\rangle$. Equation (A4) is completely equivalent to the more usual expression for potential energy matrix in the coupled basis,[44] which also separates into two $3\times 3$ blocks as follows

| $j$ | $m$ | $j = 1/2$, $m = 1/2$ | $3/2$, $1/2$ | $3/2$, $3/2$ |
|---|---|---|---|---|
| $1/2$ | $1/2$ | $(2V_\Pi + V_\Sigma)/3 - a$ | $-\sqrt{2}\,\Delta/3$ | $0$ |
| $3/2$ | $1/2$ | | $(2V_\Sigma + V_\Pi)/3 + a/2$ | $0$ |
| $3/2$ | $3/2$ | | | $V_\Pi + a/2$ |

(A5)

with an identical block for the negative $m$ states. The matrix is symmetric; for simplicity, we have not shown the lower triangle. Here we have introduced the splitting between the $\Sigma$ and $\Pi$ potentials,

$$\Delta(R) \equiv V_\Sigma(R) - V_\Pi(R) \ . \tag{A6}$$

Matrix (A5) can be further factored into a 2 × 2 block and a single state, and since no complex arithmetic is involved, the latter representation is obviously preferable for dimer complexes involving a single noble gas atom. The upper-left 2 × 2 block of matrix (A5) is identical to the $\tilde{\mathbf{V}}(R)$ matrix of Eq. (3) for the BAr complex with total angular momentum $J = 1/2$.[44] It is the eigenvalues of matrix (A5) which are shown in the lower panel of Fig. 2.

Now, suppose that instead of lying along the $z$ axis, the noble gas atom is arbitrarily positioned, with the orientation of $\mathbf{R}$ defined by the polar and azimuthal angles $(\theta, \phi)$. The matrix of the electrostatic interaction $\mathbf{V}(R)$ is transformed from the electrostatic components in Eq. (A4) by a 6 × 6 transformation matrix. This matrix consists of two identical 3 × 3 blocks, each of which is the usual matrix of the rotation $\Omega = \{\theta, \phi\}$ acting on the Cartesian unit vectors $\hat{x}$, $\hat{y}$ and $\hat{z}$,[55] namely

$$\mathbf{D}(\theta, \phi) = \begin{bmatrix} \mathbf{D}_{1/2} & 0 \\ 0 & \mathbf{D}_{-1/2} \end{bmatrix} . \tag{A7}$$

where the 3 × 3 matrix $\mathbf{D}_{1/2}$ is given by

| $q_l$ | $m_s$ | $q_l = x$, $m_s = 1/2$ | $y$, $1/2$ | $z$, $1/2$ |
|---|---|---|---|---|
| $x$ | 1/2 | $\cos\theta \cos\phi$ | $-\sin\phi$ | $\sin\theta \cos\phi$ |
| $y$ | 1/2 | $\cos\theta \sin\phi$ | $\cos\phi$ | $\sin\theta \sin\phi$ |
| $z$ | 1/2 | $-\sin\theta$ | 0 | $\cos\theta$ |

(A8)

and similarly for the matrix $\mathbf{D}_{-1/2}$ which is identical but couples the $m_s = -1/2$ states. Under this transformation, the matrix of the electrostatic interaction, in the uncoupled Cartesian basis, is given by

$$\mathbf{V}(R, \theta, \phi) = \mathbf{D}(\theta, \phi) \, \mathbf{V}(R) \, \mathbf{D}^T(\theta, \phi) \ . \tag{A9}$$

The explicit expression for the individual matrix elements of $\mathbf{V}(R, \theta, \phi)$ is identical to that given in Eq. (9) of Ref. 45, Eq. (6) of Ref. 29 , Eq. (42) of Ref. 30, and Eqs. (5) and (6) of Ref. 31.

TABLE I. B($2s^2 2p$)Ar potential energy curves (in cm$^{-1}$) used.

| R (bohr) | $V_\Pi(R)$ | $V_\Sigma(R)$ |
|---|---|---|
| 4.0 | 5773.15 | 18518.21 |
| 4.2 | 4235.05 | 14112.25 |
| 4.4 | 3024.83 | 10732.76 |
| 4.6 | 2101.58 | 8102.29 |
| 4.8 | 1418.00 | 6036.60 |
| 5.0 | 920.41 | 4444.64 |
| 5.2 | 559.47 | 3240.79 |
| 5.4 | 304.43 | 2344.96 |
| 5.6 | 129.41 | 1679.26 |
| 5.8 | 13.54 | 1184.47 |
| 6.0 | − 59.02 | 820.07 |
| 6.2 | − 102.42 | 550.20 |
| 6.4 | − 125.03 | 354.41 |
| 6.6 | − 133.68 | 216.16 |
| 6.8 | − 133.56 | 119.89 |
| 7.0 | − 127.89 | 53.22 |
| 7.2 | − 119.11 | 8.31 |
| 7.4 | − 108.83 | − 20.92 |
| 7.6 | − 98.12 | − 39.12 |
| 7.8 | − 87.59 | − 49.48 |
| 8.0 | − 77.60 | − 54.29 |
| 8.5 | − 56.35 | − 53.38 |
| 9.0 | − 40.47 | − 44.77 |
| 9.5 | − 29.11 | − 35.10 |
| 10.0 | − 21.09 | − 26.68 |
| 10.5 | − 15.45 | − 20.06 |
| 11.0 | − 11.50 | − 15.04 |
| 11.5 | − 8.66 | − 11.32 |
| 12.0 | − 6.59 | − 8.57 |
| 13.0 | − 3.91 | − 5.01 |
| 14.0 | − 2.48 | − 3.11 |
| 15.0 | − 1.61 | − 1.98 |
| 16.0 | − 1.07 | − 1.30 |

TABLE II. B($2s^23s$)Ar potential energy curve (in cm$^{-1}$) used.

| R (bohr) | $V_{3s}(R)$ | R (bohr) | $V_{3s}(R)$ |
|---|---|---|---|
| 3.43 | 22110.91 | 5.48 | – 330.08 |
| 3.45 | 14607.83 | 5.77 | – 197.89 |
| 3.47 | 9663.53 | 6.07 | – 87.88 |
| 3.48 | 7861.36 | 6.43 | 10.67 |
| 3.50 | 5180.06 | 6.61 | 53.82 |
| 3.52 | 3367.45 | 7.14 | 107.79 |
| 3.53 | 2688.84 | 7.93 | 40.33 |
| 3.55 | 1655.83 | 10.50 | – 41.46 |
| 3.57 | 934.47 | 14.00 | – 13.30 |
| 3.59 | 424.50 | 15.00 | – 9.69 |
| 3.61 | 58.24 | 16.00 | – 7.16 |
| 3.62 | – 83.31 | 17.00 | – 5.13 |
| 3.64 | – 298.21 | 18.00 | – 3.77 |
| 3.66 | – 439.80 | 19.00 | – 2.79 |
| 3.69 | – 591.24 | 20.00 | – 2.10 |
| 3.75 | – 790.08 | 21.00 | – 1.59 |
| 3.86 | – 987.99 | 22.00 | – 1.22 |
| 4.39 | – 1000.81 | 23.00 | – 0.95 |
| 4.70 | – 805.99 | 24.00 | – 0.74 |
| 4.96 | – 632.60 | 25.00 | – 0.58 |
| 5.22 | – 471.48 | | |

TABLE III. Spectroscopic constants (in cm$^{-1}$) for the observed vibrational levels of BAr($B^2\Sigma^+$) compared with theoretical predictions.[a]

| v' | Energy[b] | | $B_v$[c] | | Linewidth[d] | |
|---|---|---|---|---|---|---|
| | $^{10}$BAr | $^{11}$BAr | $^{10}$BAr | $^{11}$BAr | $^{10}$BAr | $^{11}$BAr |
| 4 | −301.7 | −325.5 | | | | |
|   | −299.9 | −323.4 | 0.335 | 0.315 | | |
| 5 | −171.9 | −198.0 | 0.311 | 0.291 | | |
|   | −170.2 | −195.3 | 0.310 | 0.293 | | |
| 6 | −59.2 | −85.1 | 0.288 | 0.272 | | |
|   | −58.1 | −83.5 | 0.285 | 0.270 | | |
| 7 | 35.7 | 10.7 | 0.258 | 0.248 | < 0.05 | < 0.05 |
|   | 36.2 | 11.8 | 0.250 | 0.247 | 0.0019 | < 0.001 |
| 8 | 101.5 | 84.6 | | | 4–6 | 2.7 |
|   | 109 | 89.5 | | | 8.8 | 2.0 |
| 9 | | 128 | | | | 29 |
|   | | 143 | | | | 42 |
| $D_0$ | 994 ± 9[e] | 999 ± 9[e] | | | | |
|   | 1001.3 | 1005.4 | | | | |

a. The upper and lower entries are the experimentally determined and theoretically calculated values, respectively.
b. The zero of energy is defined as the B($2s^2 3s\ ^2S$) + Ar asymptote. The experimental energies were calculated from the transition wavenumbers $T'(v')$ presented in Table II of Ref. 3 as follows: $E(v') = T'(v') - \hbar\omega(B\ ^2S \leftarrow\ ^2P_{1/2}) - D_0''$, where $\hbar\omega(B\ ^2S \leftarrow\ ^2P_{1/2}) = 40039.7$ cm$^{-1}$ (Ref. 15) and $D_0''(^{11}BAr) = 102.4 \pm 0.3$ cm$^{-1}$ (Ref. 40). The $^{11,10}$B isotope shift for BAr($X$, v''=0) was calculated to be 0.88 cm$^{-1}$ (Ref. 3). The estimated uncertainties in the experimental energies are ± 0.5 cm$^{-1}$.
c. Rotational constant. Experimental values taken from Ref. 3.
d. Linewidth of predissociating levels. The theoretical values were obtained by fitting the square of the calculated bound-free transition moment to a Lorentzian (see Sec. III for more details).
e. Calculated from $E(v' = 4)$ and the mass-scaled vibrational parameters $\omega_e'$ and $\omega_e x_e'$ reported in Ref. 3.

TABLE IV. Morse parameters for vibrational ground state of the BAr and Ar$_2$ complexes.

| parameter | BAr($X^2\Pi_{1/2}$) | Ar$_2$ |
|---|---|---|
| $D_e$ (cm$^{-1}$) | 125.0 | 97.811 |
| $r_e$ (bohr) | 6.6889 | 7.1053 |
| $\beta$ (bohr$^{-1}$) | 0.83451 | 0.89305 |

FIGURE CAPTIONS

FIG 1. Survey laser fluorescence excitation scan, taken with 9.1 atm source backing pressure, showing the higher $B^2\Sigma^+$–$X^2\Pi$ (v′,0) bands for the $^{11,10}$BAr isotopomers. Also marked are the B atomic $2s^23s\ ^2S - 2s^22p\ ^2P_{1/2,3/2}$ lines and the feature at high energy arising from the $3s \leftarrow 2p$ transition in the ternary BAr$_2$ complex. The fluorescence signal has not been corrected for the small wavelength variation of the laser pulse energy. The atomic B and BAr features are identical to those shown in Fig. 1 of Ref. 3. The abscissa is the shift (in cm$^{-1}$) relative to the B($^2S \leftarrow\ ^2P_{1/2}$) transition [$h\nu$ = 40039.7 cm$^{-1}$ (Ref. 15)].

FIG 2. Potential energy curves for the BAr $X^2\Pi_{1/2}$, $A^2\Sigma^+$, and $B^2\Sigma^+$ states. The $X$ and $A$ state curves correspond to our earlier *ab initio* calculations (Ref. 3) modified by the scaled correlation energy adjustment described in Sec. III with a scaling factor of $s = 1.19$. The $B$ state curve is the RKR curve from Ref. 3 with the position of the vibrational energy levels adjusted with the help of our recent determination of the ground dissociation energy, as described in the text. For all states, the asymptotes refer to the separated atom limits [B(2p, 3s) + Ar]. The position, height, and shape of the barrier (shown in more detail in the inset) have been adjusted to give reasonable agreement (see Table III) with the experimental estimates of the positions and widths of the predissociating v′ = 8 and 9 levels. In the inset, the upper and lower lines, for each value of v, indicate the position and outer turning point of the vibrational levels for $^{10}$BAr and $^{11}$BAr, respectively. In the case of v′ = 9, the resonance associated with the lighter isotope is too broad to be either seen in the experimental spectrum or distinguished in the theoretical simulation of the spectrum.

FIG 3. (Upper panel) Comparison of the variational energy [Eq. (12)] of the lowest vibrational state of the BAr$_2$ complex, calculated using the variational Monte-Carlo method within the electronically adiabatic approximation. The three traces correspond to $N$ = {100, 500, 500} integration points and initial widths of $dq$ = {4, 4, 5}, respectively. For clarity, open and filled circles are used to mark the traces corresponding to the second and third sets of parameters. The root-mean-square energies and standard deviations for the three sets of parameters, determined using the calculated energies from iterations 61 – 80, are (in

cm$^{-1}$), $E_{rms}$ = {–295.08, –294.91, –295.00} and $\sigma$ = { 0.51, 0.22, 0.21}. The dashed horizontal line corresponds to an energy of –295.1 cm$^{-1}$. (Lower panel) Comparison of the DMC reference energy as a function of imaginary time for the lowest vibrational state of the BAr$_2$ complex. An ensemble of 2000 replicas was used with a time step of 0.25 a.u. The solid line and open circles indicate the energies determined without and with, respectively, importance sampling based on the variational approximation of Eq. (14) discussed in Sec. IVA. The dashed horizontal line corresponds to an energy of –295.9 cm$^{-1}$.

FIG 4. Potential energy surface for the B(3$s$)Ar$_2$ complex as a function of the two BAr distances. The ArBAr angle is 64.16°, corresponding to the angle at the minimum in the B(2$p$)Ar$_2$ cluster. All energies are in cm$^{-1}$, with respect to B(3$s$)+Ar+Ar. For clarity, the zero energy contours are marked with heavy lines. The dashed contours represent the square of the lowest vibrational wavefunction of the B(2$p$)Ar$_2$ complex.

FIG 5. Comparison of the predicted laser excited fluorescence spectrum (top and middle panel) of the BAr$_2$ complex with the experimental observations (lower panel). The abscissa is the shift (in cm$^{-1}$) relative to the B($^2S \leftarrow\ ^2P_{1/2}$) transition [$h\nu$ = 40039.7 cm$^{-1}$ (Ref. 15)]. The upper panels presents spectral simulations carried out with points and weights resulting from application of the DMC method (upper panel) and VMC method (middle panel). The solid and dashed spectra correspond, respectively to application of the semiclassical expression with [Eq. (30)] and without [Eq. (28)] correction for the kinetic energy of the B atom. The lower panel (experimental spectrum) is an enlargement of the BAr$_2$ feature shown in Fig. 1. As indicated by the dashed curve in the lower panel, and seen more clearly in Fig. 1, at energies less than ~ 290 cm$^{-1}$, this feature is obscured by the wing of the BAr ($B^2\Sigma^+$, v' = 9) $\leftarrow$ ($X^2\Pi_{1/2}$, v" = 0) transition, which is centered at 230 cm$^{-1}$ and substantially broadened by predissociation.

Fig. 6. Contour plots of the potential difference $V_{3s}(q_i) - V_{2p}(q_i)$ (solid contours) and the square of the ground state vibrational wavefunction $|\Phi_g(q)|^2$ (dashed contours), for BAr$_2$. As in Fig. 4, the two BAr distances are varied, but the ArBAr angle is held to 64.16°, corresponding to the angle at the minimum in the B(2$p$)Ar$_2$ cluster. All energies

are in cm$^{-1}$, with respect to B(3$s$)+Ar+Ar. In innermost energy contour is 450 cm$^{-1}$. Contours for $|\Phi_g(\boldsymbol{q})|^2$ are plotted at {0.8, 0.6, 0.4, and 0.2}.

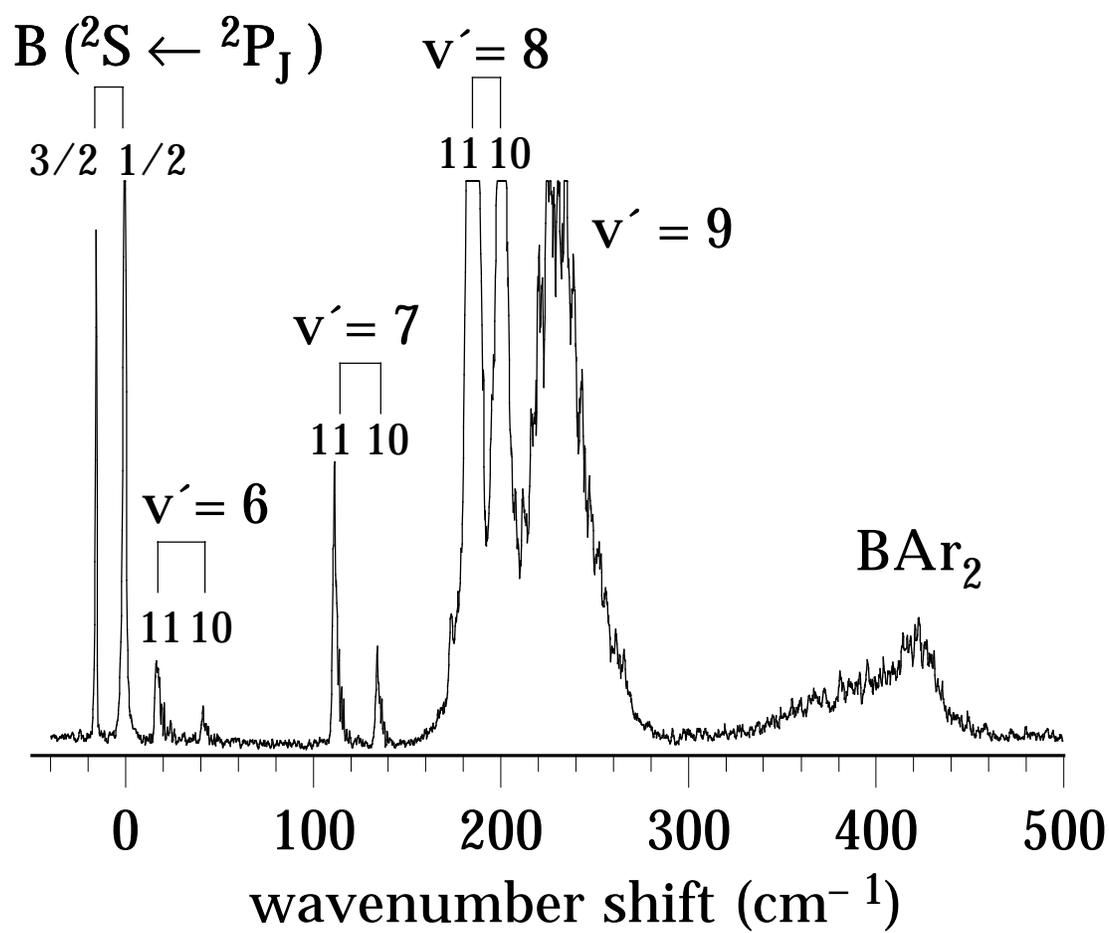

Fig. 1. M. H. Alexander et al. "A collaborative theoretical and experimental study of the structure and electronic excitation spectrum of ...," J. Chem. Phys.

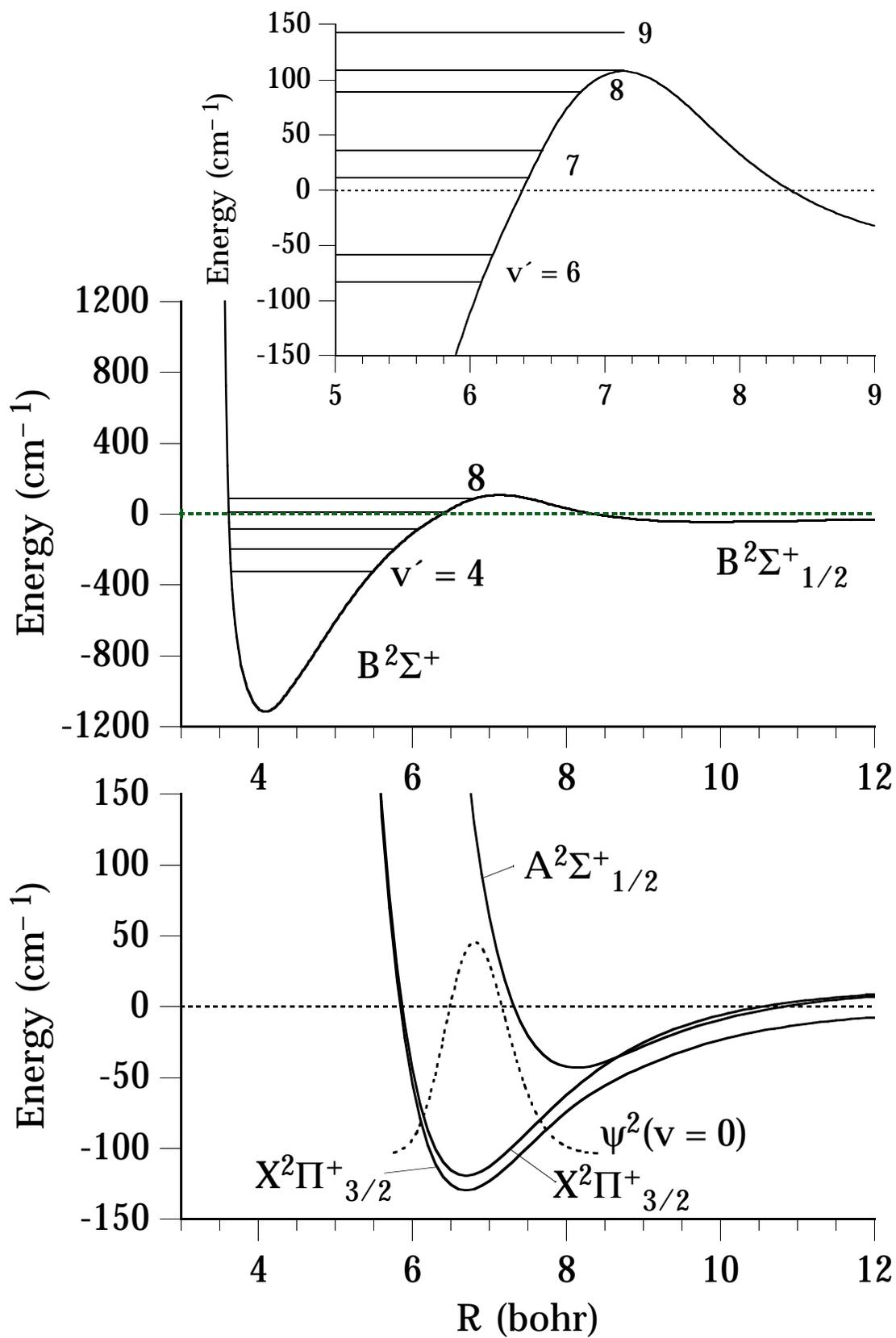

Fig. 2. M. H. Alexander et al., "A collaborative theoretical and experimental study of the structure and electronic excitation spectrum ...," J. Chem. Phys.

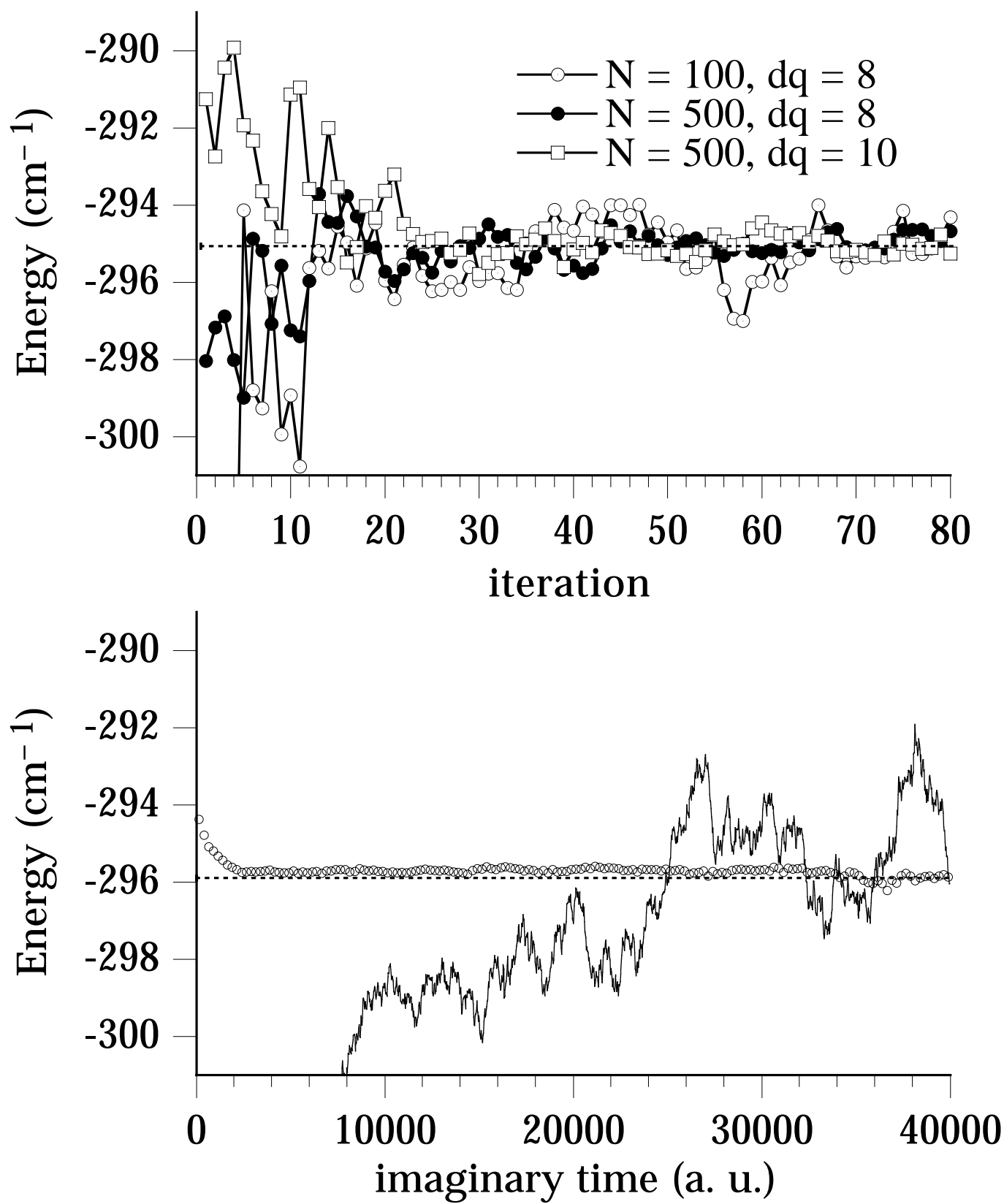

Fig. 3. M. H. Alexander et al. "A collaborative theoretical and experimental study of the structure and electronic excitation spectrum of ...," J. Chem. Phys.

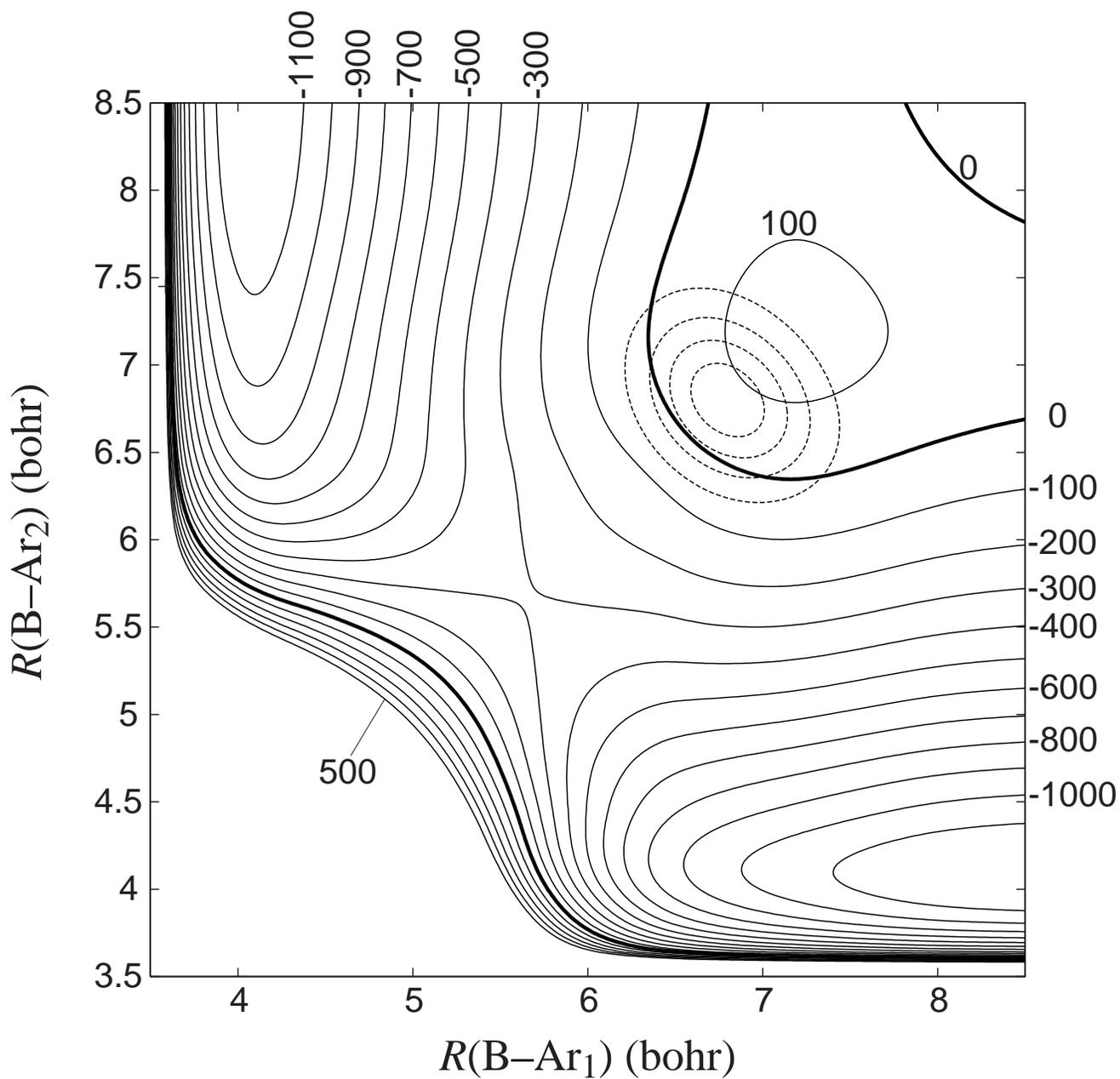

Fig. 4. M. H. Alexander *et al.*, "A collaborative theoretical and experimental study of the structure and electronic excitation spectrum of the BAr(*X, B*) and B(Ar)$_2$ complexes," J. Chem. Phys.

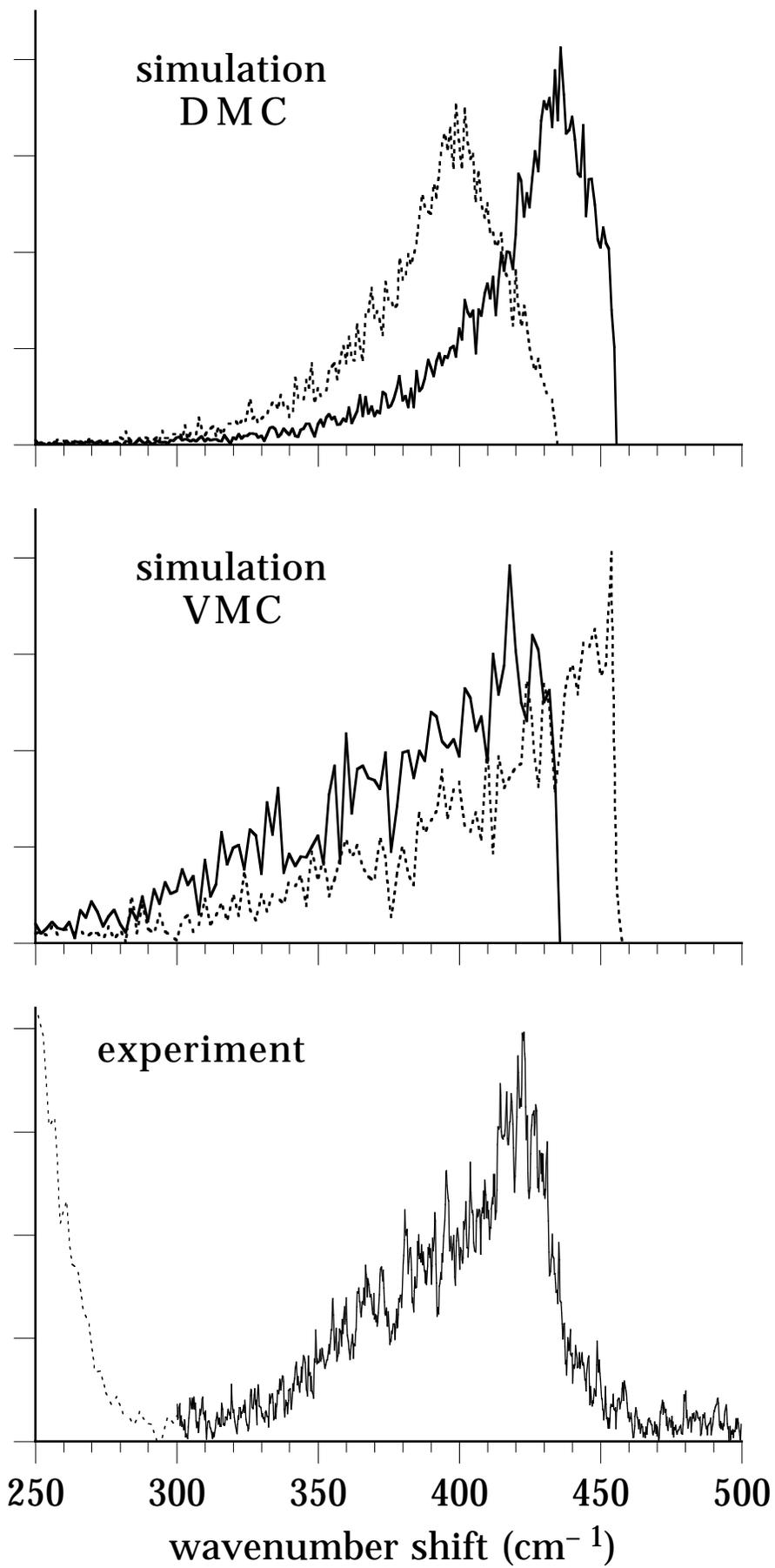

Fig. 5. M. H. Alexander et al. "Combined theoretical and experimental investigation of the BAr$_2$ complex," J. Chem. Phys.

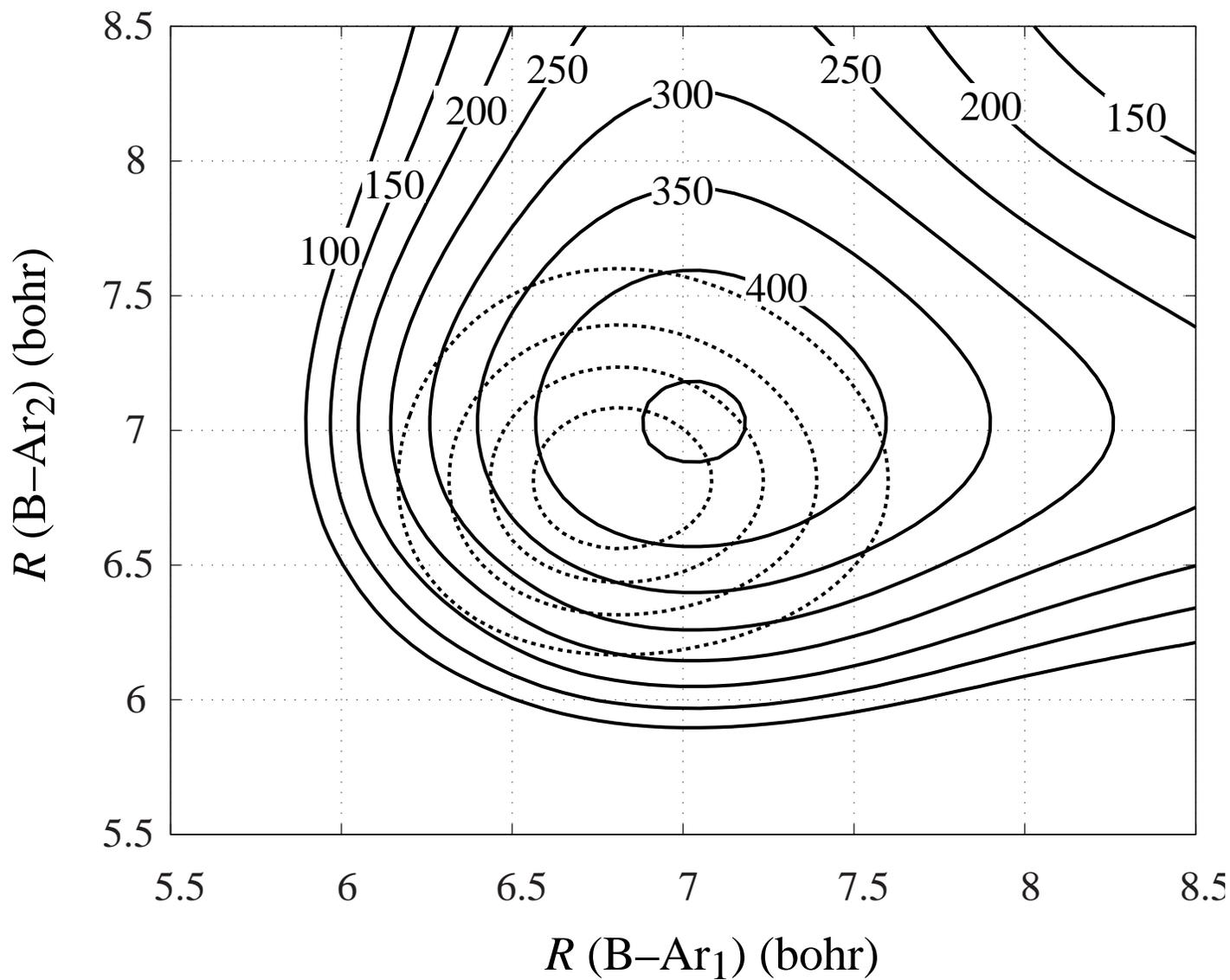

Fig. 6. M. H. Alexander *et al.*, "A collaborative theoretical and experimental study of the structure and electronic excitation spectrum of the BAr(*X, B*) and B(Ar)$_2$ complexes," J. Chem. Phys.